# Enhancing atomic-resolution in electron microscopy: A frequency-domain deep learning denoiser

Ivan Pinto-Huguet[1,*,†], Marc Botifoll[1,†], Xuli Chen[1], Martin Børstad Eriksen[2], Jing Yu[1,3], Giovanni Isella[5], Andreu Cabot[3,4], Gonzalo Merino[2], and Jordi Arbiol[1,4,*]

[1]Catalan Institute of Nanoscience and Nanotechnology (ICN2), CSIC and BIST, Campus UAB, Bellaterra, 08193, Barcelona, Catalonia, Spain
[2]PIC, IFAE, Campus UAB, Bellaterra, 08193, Barcelona, Catalonia, Spain
[3]Catalonia Institute for Energy Research (IREC), Barcelona 08930, Spain
[4]ICREA, Passeig Lluís Companys 23, 08010, Barcelona, Catalonia, Spain
[5]Laboratory for Epitaxial Nanostructures on Silicon and Spintronics, Physics Department, Politecnico di Milano, Como, Italy

*Corresponding authors: ivan.pinto@icn2.cat, arbiol@icrea.cat;
†Equal contribution

**Abstract**

The study of functional materials increasingly demands atomic-scale analysis to explore their fundamental properties, requiring advanced imaging techniques capable of resolving individual atoms and accurate atomic column positions in the structures. Atomic resolution electron microscopy, particularly high-angle annular dark-field scanning transmission electron microscopy, has become an essential tool for many scientific fields, when direct visualization of atomic arrangements and defects are needed, as they dictate the material's functional and mechanical behavior. However, achieving this precision is often hindered by noise, arising from electron microscopy acquisition limitations, particularly when imaging beam-sensitive materials or light atoms. In this work, we present a deep learning-based denoising approach that operates in the frequency domain using a the convolutional neural network U-Net trained on simulated data. To generate the training dataset, we simulate FFT patterns for various materials, crystallographic orientations, and imaging conditions, introducing noise and drift artifacts to accurately mimic experimental scenarios. The model is trained to identify relevant frequency components, which are then used to enhance experimental images by applying element-wise multiplication in the frequency domain. The model enhances experimental images by identifying and amplifying relevant frequency components, significantly improving signal-to-noise ratio while preserving structural integrity. Applied to both Ge quantum wells and $WS_2$ monolayers, the method facilitates more accurate strain quantitative analyses, critical for assessing functional device performance (e.g. quantum properties in SiGe quantum wells), and enables the clear identification of light atoms in beam sensitive materials. Our results demonstrate the potential of automated frequency-based deep learning denoising as a useful tool for atomic-resolution nanomaterials analysis.

# 1 Introduction

Transmission electron microscopy (TEM) is a powerful tool for studying materials at the atomic scale, providing critical insights into structural, functional, and chemical properties. [1, 2, 3, 4]However, experimental images often suffer from noise introduced by various sources, such as beam fluctuations, sample drift, and detector limitations. [5, 6] This noise significantly affects the visibility of atomic structures, particularly in low-contrast materials, making reliable image interpretation challenging. Traditional denoising techniques, such as Gaussian filtering, wavelet transforms, and Wiener filtering, can mitigate noise to some extent but often require manual parameter tuning and may introduce artifacts or degrade high-frequency information. [7, 8, 9] Another commonly used approach to mitigate this noise is Fast Fourier Transform (FFT) filtering, which isolates frequency components related to periodic atomic arrangements. By selectively removing noise-related frequencies, FFT filtering can enhance the clarity of crystallographic features and improve the study of crystal properties. [10] However, the process of identifying the relevant frequencies is not always straightforward. In experimental data, this frequency selection is often performed manually, which introduces a degree of human bias and subjectivity. Moreover, the very act of filtering can inadvertently remove critical information, leading to the introduction of artifacts and compromising the accuracy of the analysis. These limitations make traditional FFT-based denoising prone to errors, especially when dealing with real-world experimental conditions.

In recent years, deep learning-based methods, particularly Convolutional Neural Networks (CNNs), have shown great success in data processing for TEM applications. [11, 12] Among the various CNN architectures used for image processing tasks, U-Net stands out. [13, 14, 15] Originally designed for biomedical image segmentation, U-Net has been widely adopted in scientific imaging due to its ability to preserve fine details while suppressing noise. [16] CNNs, in particular, have demonstrated significant promise for denoising in real space . [14, 15, 17, 18] However, their application in reciprocal space, where crucial information about crystal structures resides, has yet to be thoroughly explored. Existing CNN-based denoising models predominantly operate in real space, leaving untapped potential for substantial improvements in the reciprocal space domain.

In this work, we propose a deep learning-based FFT denoising approach using a U-Net model trained on simulated data. These simulations encompass multiple materials, crystallographic orientations, and imaging conditions, ensuring that the trained model learns to identify meaningful frequency components across a wide range of experimental scenarios. By introducing controlled noise and drift artifacts, we create a dataset that accurately mimics real imaging conditions, enabling the network to generalize effectively to experimental data.[5, 6, 19]

The proposed denoising pipeline consists of three key steps. First, the experimental FFT is passed through the trained U-Net model, which identifies the locations of the meaningful frequency components. Next, the predicted frequency map is used to filter the original FFT via element-wise multiplication, resulting in what we define as the denoised FFT. Finally, the inverse FFT (IFFT) is applied to reconstruct a denoised real-space image with an improved signal-to-noise ratio (SNR). This approach not only enhances the visibility of atomic features but also preserves essential structural information, surpassing the limitations of traditional FFT filtering methods. Unlike conventional techniques that mask or discard specific frequencies, our model enhances the relevant frequencies, removing noise while retaining critical crystallographic details. By eliminating the subjective human bias in frequency selection and avoiding the introduction of artifacts, our method provides a more precise and reliable solution for electron microscopy image denoising.

To validate our model, we applied it to HAADF-STEM images of two technologically relevant material systems: Ge quantum wells (QWs) embedded in SiGe barriers and a 2D $WS_2$ monolayer. [3, 20] SiGe-based heterostructures are essential in advanced semiconductor technology, particularly for quantum computing and optoelectronic applications. Ge QWs embedded in SiGe barriers serve as an ideal benchmark due to the similar lattice parameters between the Ge QWs and the SiGe barriers, which have a high stoichiometric Ge content (around 80%). Precise strain measurements at the QWs are necessary to understand their behavior and to model the final device properties accurately. The close cell parameter values make it nearly impossible to accurately measure the strain, especially when the atomic resolution HAADF STEM images are noisy. The characteristic dumbbell-shaped atomic columns in the SiGe lattice, when visualized along the [110]-axis, need to be clearly resolved for accurate structural analysis. Conventional denoising methods often blur these fine details, whereas our model enhances their visibility by selectively amplifying relevant frequency components while suppressing noise. The application of Geometric Phase Analysis (GPA) on both denoised and raw images demonstrates that our denoised strain maps have significantly higher homogeneity, allowing for more reliable strain measurements, which are crucial for evaluating the performance of these quantum devices. On the other hand, $WS_2$ is a prototypical transition metal dichalcogenide (TMD) widely explored for next-generation electronic and optoelectronic devices. In this case, the challenge lies in its weak scattering contrast, particularly for sulfur atoms, which are often difficult to distinguish in noisy experimental HAADF-STEM images. Traditional filtering or real space denoising techniques may either suppress sulfur signal or introduce artifacts, while our approach effectively enhances its visibility, enabling a more accurate representation of the atomic structure. These results demonstrate the robustness of our model across different material systems and highlight its potential for improving atomic-scale analysis in electron microscopy.

## 2 Methods

To develop a robust denoising methodology of (S)TEM images, we employed a CNN based on the U-Net architecture. The model was trained to denoise the reciprocal space representation of the images, which inherently improves the real-space images as well. This section describes the simulation pipeline used to generate training data, the network training process, and the implementation of the denoising strategy in real-space images using its respective denoised FFT.

### 2.1 Simulating the training dataset

To train the model, we require a large dataset of experimental FFTs with their corresponding targets/labels. In this case, the labels consist of an image where the frequency coordinates are assigned a value of 1, while the rest are set to 0, representing the ideal denoised FFT. Due to the lack of experimentally labeled FFTs and the time-consuming nature of manually labeling a sufficiently large dataset, our approach was to simulate these FFTs for a collection of materials, zone axes, and field of views (FOVs). These simulations are based on kinematic diffraction theory to generate 2D diffraction patterns corresponding to the given crystal phase model. [21].

The simulated dataset includes seven different materials: Ge (diamond cubic), GaAs (wurtzite), InP (zinc-blende), InSb (zinc-blende), Si (diamond cubic), $WS_2$ (graphene), and ZnO (wurtzite). These materials were chosen for their comparable yet distinct atomic structures. Simulations were performed for orientations along major crystallographic axes ([001], [100], [101], [110], and [111]) and at three different resolutions: 0.015 nm/pixel, 0.03 nm/pixel, and 0.05 nm/pixel, resulting in more than 100 unique simulation conditions. The selection of different materials and orientations ensured the robustness of the model, allowing it to generalize across various experimental conditions. Additionally, variations in pixel size improved adaptability to different electron microscopy setups.

Since these simulations generate a theoretical diffraction pattern, they can serve as labels, as they exclusively contain the frequency positions. However, to train the model, we still need the experimental-like FFTs, to serve as our input data. To generate these, we attempted to mimic experimental conditions by applying different layers of noise and drift to the simulated FFTs. The following paragraphs describe, step by step, the process of transforming a simulated FFT into an experimental-like noisy FFT (Fig. 1), as used in this work.

The first transformation introduces a drift artifact in the simulated FFT to emulate sample movement during acquisition (Fig. 1b). This involves displacing the frequencies by a horizontal offset $s$ and rotating the image by an angle $\theta$ (S1.1). This step not only replicates real experimental conditions but also ensures that the dataset encompasses a wide range of potential deviations, breaking symmetry patterns and reducing the risk of overfitting. After this transformation, we apply an inverse FFT to the drifted FFT to transition into real space, as all subsequent noise filters are applied in this domain.

As seen in Fig. 1c, since the FFT originates from discrete frequency points, the atoms in the corresponding real-space image also appear as punctual-like structures. To mimic experimental atomic images, we apply a Gaussian filter that provides a two-dimensional shape to the atoms (Fig. 1d). By modulating the $\sigma$ parameter of the Gaussian distribution, we control the atomic radius, simulating the scattering profiles observed in experimental images (S1.2). Following this transformation, additional noise layers are applied to mimic scanning aberrations and detector noise.

The first noise layer simulates scanning noise by introducing Gaussian distributed displacements in both the vertical and horizontal directions (S1.3). This noise is controlled by the variance parameters $\sigma_h$ and $\sigma_v$. The second noise layer is the shot noise, also known as Poisson noise (S1.4), regulated by the $\lambda$ parameter of the Poisson distribution. The combined effect of these two noise layers results in an atomic image that closely resembles an experimental one (Fig. 1e). Finally, we compute the FFT of the noisy image to obtain the final noisy FFT, which mimics an experimental FFT (Fig. 1f).

For each of the 100 different simulated FFTs, all the transformations described above were applied to generate the training dataset. Each simulation produced 50 different noisy FFTs, regulated by the parameters introduced earlier: the drift parameters $s$ and $\theta$, the Gaussian filter parameter $\sigma$, the scanning noise variances $\sigma_h$ and $\sigma_v$, and the Poisson noise parameter $\lambda$. These parameters were randomly sampled within predefined ranges (S1.5), ensuring diverse noise levels and drift conditions for each simulated FFT. As a result, the final dataset consisted of over 5000 FFT spectra.

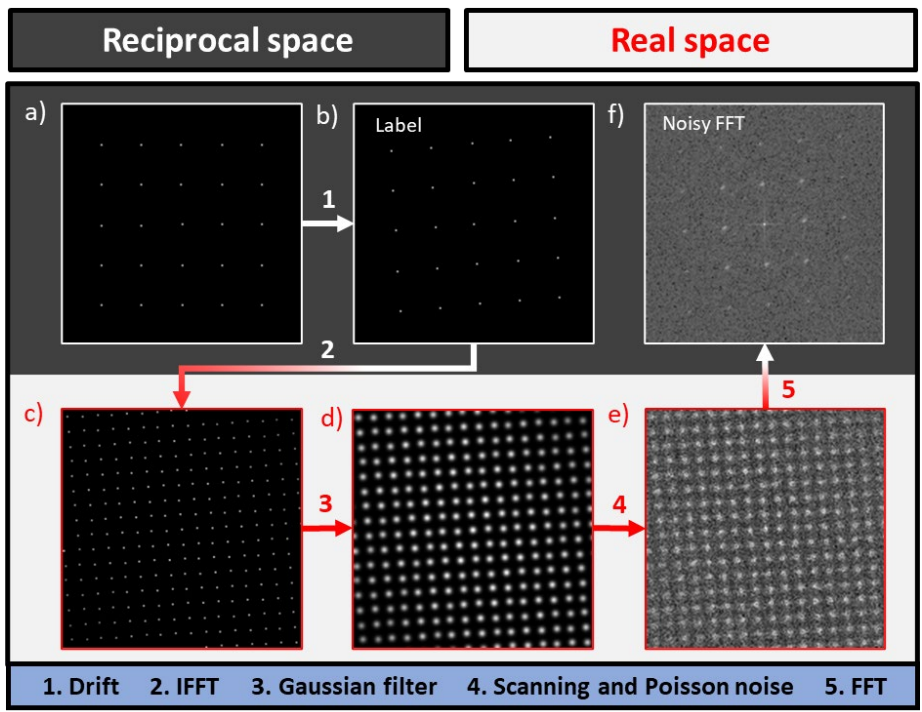

Figure 1: **Pipeline for generating a noisy Fast Fourier Transform (FFT) spectrum from a simulated FFT.** (a) Simulated FFT. (b) Drifted FFT. (c) Image of punctual-like atoms. (d) Image of Gaussian atoms. (e) Atomic image with noise mimicking experimental conditions. (f) Noisy FFT, representing an experimental-like FFT. The spectra in (b) and (f), highlighted in red, are used for training: (f) serves as input data, while (b) is used as the label.

## 2.2 U-Net training

For model training, the noisy FFT was used as the input data, while a modified version of the drifted FFT served as the label (Figs. 1f and 1b, respectively). This modification involved transforming single-pixel frequency peaks into small circular regions, as shown in Fig. 2a. This adjustment served two purposes. First, it improved model convergence, as using single-pixel labels could lead to drastic errors—where a one-pixel shift in prediction could mean the difference between a perfect and a completely incorrect output. Second, in experimental FFTs, frequency information is typically spread across multiple pixels rather than concentrated in a single point. Therefore, detecting a small region around the frequency peak provided a more realistic approach.

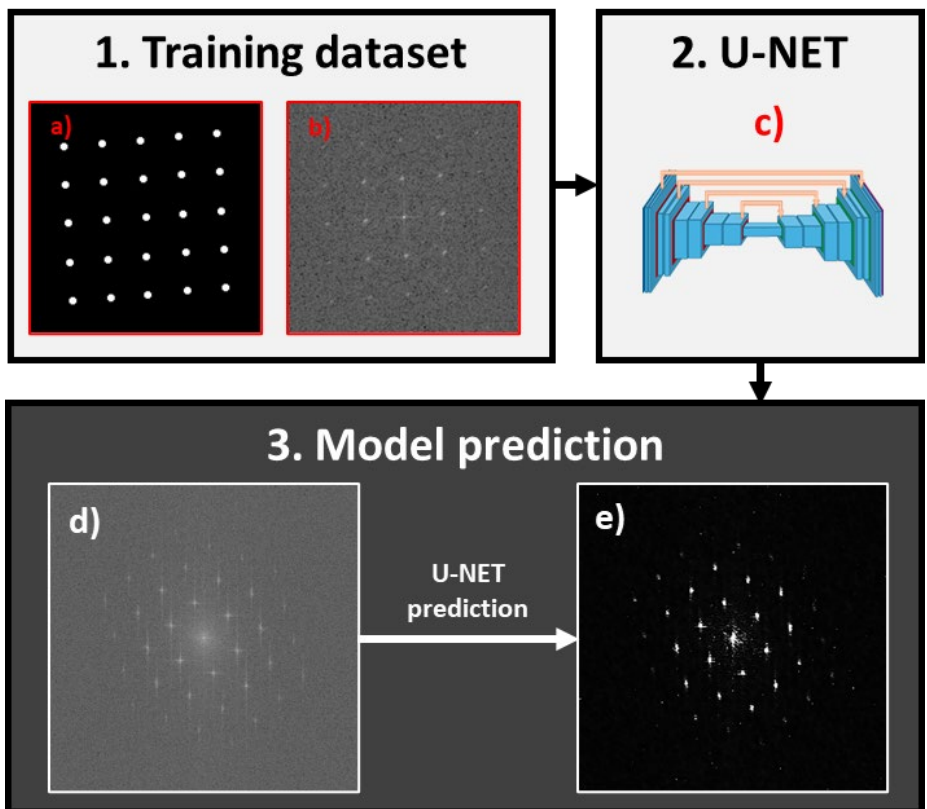

Figure 2: (a) Circular frequency labels and (b) noisy FFT used as label and input data, respectively, for training the Convolutional Neural Network (CNN) model. (c) Schematic of the U-Net architecture, which follows an encoder-decoder structure. (d) Experimental FFT used for testing. (e) U-Net prediction on the experimental FFT after training.

We used the dataset of 5000 FFT pairs, like those shown in Figs. 2a and 2b, to train a U-Net model. The U-Net architecture was implemented using the AtomAI Python package [22]. The loss function was cross-entropy loss, which is well-suited for binary classification tasks. To minimize this loss, we used Stochastic Gradient Descent (SGD) as the optimizer and enabled Stochastic Weight Averaging to stabilize training and improve generalization. The model was trained for 500 epochs with a batch size of 32 and a learning rate of 0.001.

After training, the model achieved an accuracy of 99.8% on the training data and 99.4% on the test data, indicating high performance in the simulated dataset. However, such high accuracy could suggest potential overfitting, meaning the model might struggle to generalize to experimental data. To evaluate its performance under real conditions, we tested the model on an experimental FFT from a $WS_2$ monolayer sample (Fig. 2d). The model produced satisfactory results, as seen in Fig. 2e, accurately detecting the low-frequency regions of interest and identifying higher-frequency components, although with lower intensity.

## 2.3 Denoising in real space: Global algorithm

The general algorithm utilizes the trained U-Net model to denoise experimental images. In the following paragraphs, we describe the entire denoising process. As an example,we use an experimental image of a Ge QW in a SiGe heterostructure cross-section to illustrate each step (Fig. 3).

First, we compute the FFT of the image and use the U-Net model to predict the regions corresponding to the most significant frequencies. Next, we apply element-wise multiplication (Hadamard product) to obtain the denoised FFT. The key idea behind this step is that the U-Net provides a frequency map highlighting the relevant regions. By multiplying this output with the original FFT, we enhance the signal from crystalline regions, particularly improving atomic contrast. As shown in Figure 3e, the intensity difference between the background and the frequency peaks is doubled in the denoised FFT compared to the original.

Finally, we compute the IFFT of the denoised FFT to reconstruct the denoised image. As seen in Figure 3f, this process significantly improves the SNR of the original image. In the next section, we will discuss these results in detail.

## 3 Results: beyond the limits of HAADF-STEM imaging

In this section, we demonstrate the effectiveness of our denoising algorithm in two distinct scenarios: enhancing contrast in highly noisy conditions and revealing hidden atomic features. First, we apply our method to a HAADF-STEM image of a Ge QW embedded in a SiGe heterostructure , where we quantify the improvement in signal-to-noise ratio (SNR) and assess the enhanced visibility of atomic dumbbells, which are unresolved in the raw image. Additionally, we perform GPA on the denoised image to extract strain maps, showing a significant reduction in noise and more accurate and homogeneous strain measurements, which are critical for evaluating quantum device performance. [20, 23, 24] Next, we extend our analysis to a monolayer $WS_2$ sample, where the inherent HAADF-SETM contrast difference between W and S atoms makes sulfur nearly undetectable in the original image. Through controlled simulations, we confirm that our algorithm reveals real structural features by suppressing background noise. Finally, we compare our model with a real-space denoising approach, and demonstrate that only our frequency-based model can successfully show the presence of sulfur atoms, highlighting its superior capability in revealing hidden atomic features.

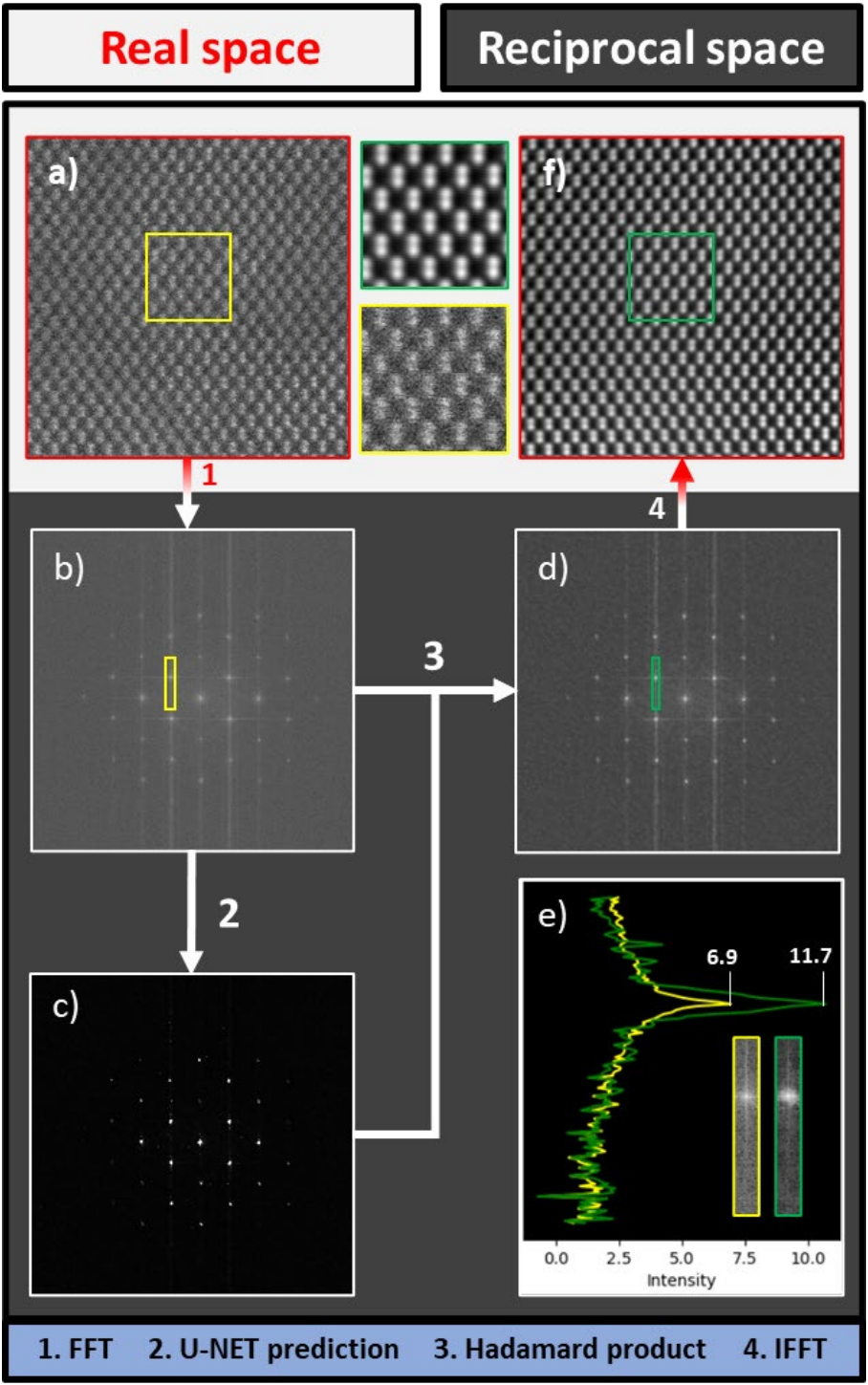

Figure 3: **Global pipeline of the denoising algorithm in real space.** (a) Experimental HAADF-STEM image of a selected region of interest in a Ge QW cross-section. (b) Fast Fourier Transform (FFT) spectrum of (a). (c) U-Net prediction of (b) after training on simulated data. (d) Denoised FFT obtained by applying the Hadamard product to (b) and (c). (e) Intensity comparison of a frequency peak in the original FFT (yellow) and the denoised FFT (green), showing signal enhancement. (f) Inverse FFT of (d), yielding the final denoised image. The images in (a) and (f) present a field of view (FOV) of 7.5 nm.

## 3.1 Enhancing relative strain mapping and atomic resolution in Ge QW images

To demonstrate the effectiveness of our denoising algorithm, we apply it to a HAADF-STEM image of a Ge QW cross-section (Fig. 4e).The image corresponds to a SiGe heterostructure composed of central pure Ge QW embedded between $Si_{0.2}Ge_{0.8}$ barriers. The full heterostructure presents a diamond cubic crystal phase (space group 227). The heterostructure is grown along the [001] direction and it is visualized along its [110] zone axis.

To quantitatively assess the denoising performance, we use the signal-to-noise ratio (SNR) as a metric. The SNR is calculated as the ratio between the maximum intensity in the atomic regions and the average intensity of the vacuum background. In the raw image (Fig. 4a), we obtain an SNR of 2, whereas in the denoised image (Fig. 4b), the SNR increases to 10. This demonstrates that our algorithm enhances the SNR by up to a factor of five . However, the benefits of the denoising process extend beyond SNR improvement. As seen in Figures4b and 4c, the denoised image exhibits significantly higher contrast and sharpness. Furthermore, the intensity profile comparison between the raw and denoised images (Fig. 4c) reveals a smoother shape and a greater difference between maxima and minima, in agreement with the improved SNR.

A key outcome of our denoising approach is the improved resolution of atomic features. In the original image, the Ge dumbbells are not clearly resolved due to noise, whereas in the denoised image, individual atoms within the dumbbell structure become clearly resolvable. This is further evidenced by the intensity profiles across the dumbbells in Fig. 4d. The profile from the raw image (yellow) shows a broad, unresolved peak, whereas the denoised image (green) clearly reveals a double peak, corresponding to the two individual atoms in the dumbbell. This result closely matches the theoretical dumbbell profile (blue), highlighting the algorithm's ability to recover fine structural details that are otherwise lost in noisy conditions.

Beyond structural visualization, the enhanced image quality also enables more accurate quantitative analysis. We performed GPA to extract relative strain maps across the quantum well, a critical parameter in the performance of quantum devices. The GPA results obtained from the denoised images exhibit clearer and more interpretable strain distributions compared to those derived from the raw data (Figs. 4g and 4h).

To quantitatively assess the improvement, we calculated the average relative strain values in the bottom, middle, and top layers of the SiGe heterostructure, corresponding to the $Si_{0.2}Ge_{0.8}$ bottom barrier, the Ge QW and the $Si_{0.2}Ge_{0.8}$ top barrier, respectively . In the original (noisy) $\epsilon_{yy}$ strain map, we obtained values of 0.0 ± 5.1%, 1.6 ± 4.4% and −0.8 ± 5.7% respectively. After applying our denoising algorithm, these $\epsilon_{yy}$ values were refined to 0.0 ± 0.6%, 1.3 ± 0.7% and −0.9 ± 0.5%. These results demonstrate a dramatic reduction of noise (standard deviation), by approximately one order of magnitude, while preserving the physical validity of the strain values. Notably, the experimental $\epsilon_{yy}$ strain values from the raw map lie within the confidence intervals of the denoised results, confirming that the algorithm enhances signal fidelity without introducing artifacts. Additional strain components relevant for GPA analysis are included in Table S2 and Figure S5, further illustrating the robustness and consistency of our denoising method across different strain directions. This improvement is particularly relevant since even minimal strain variations at the nanoscale can strongly influence the electronic band structure and, consequently, the functionality of quantum wells and related devices. Our denoising framework not only improves structural clarity but also unlocks precise strain quantification, enabling robust strain engineering strategies crucial for the development of future quantum technologies.

## 3.2 Resolving hidden features: light atoms visualization in WS$_2$ monolayers

In this case, we demonstrate that our algorithm can reveal features that are otherwise obscured by noise. To validate this capability, we use a monolayer HAADF-STEM image of a 2D material as an example. The selected image corresponds to WS$_2$ in its hexagonal 2H phase (space group 194) visualized along its basal [0001] zone axis (Fig. 5).

In HAADF-STEM mode, the intensity is approximately proportional to the atomic number squared ($\approx Z^2$). This results in a significant contrast difference between W and S atoms, where the intensity of S is often hidden in the background noise, as observed in Fig.5i. After applying our denoising algorithm, we not only observe an improvement in the SNR of W atoms (from 1.5 to 5.5) but also the appearance of low-intensity peaks at the expected positions of sulfur atoms (Fig.5l). This raises two possibilities: either the algorithm introduces periodic artifacts due to modifications in the reciprocal space, or it successfully unveils hidden structural features by suppressing background noise.

To distinguish between these two scenarios, we conducted a controlled simulation test. We generated two simulated HAADF-STEM images of WS$_2$ one including both W and S atoms, and another where S atoms were removed . Noise was added to both images to mimic experimental conditions, ensuring that S intensities remained indistinguishable in the raw data. Finally, we applied our denoising algorithm to both cases. If both simulations exhibited intensity peaks at the S positions after denoising, it would indicate that the algorithm introduces periodic artifacts. However, if only the image that originally contained S atoms displayed these intensity peaks, it would confirm that the algorithm can retrieve hidden existing features.

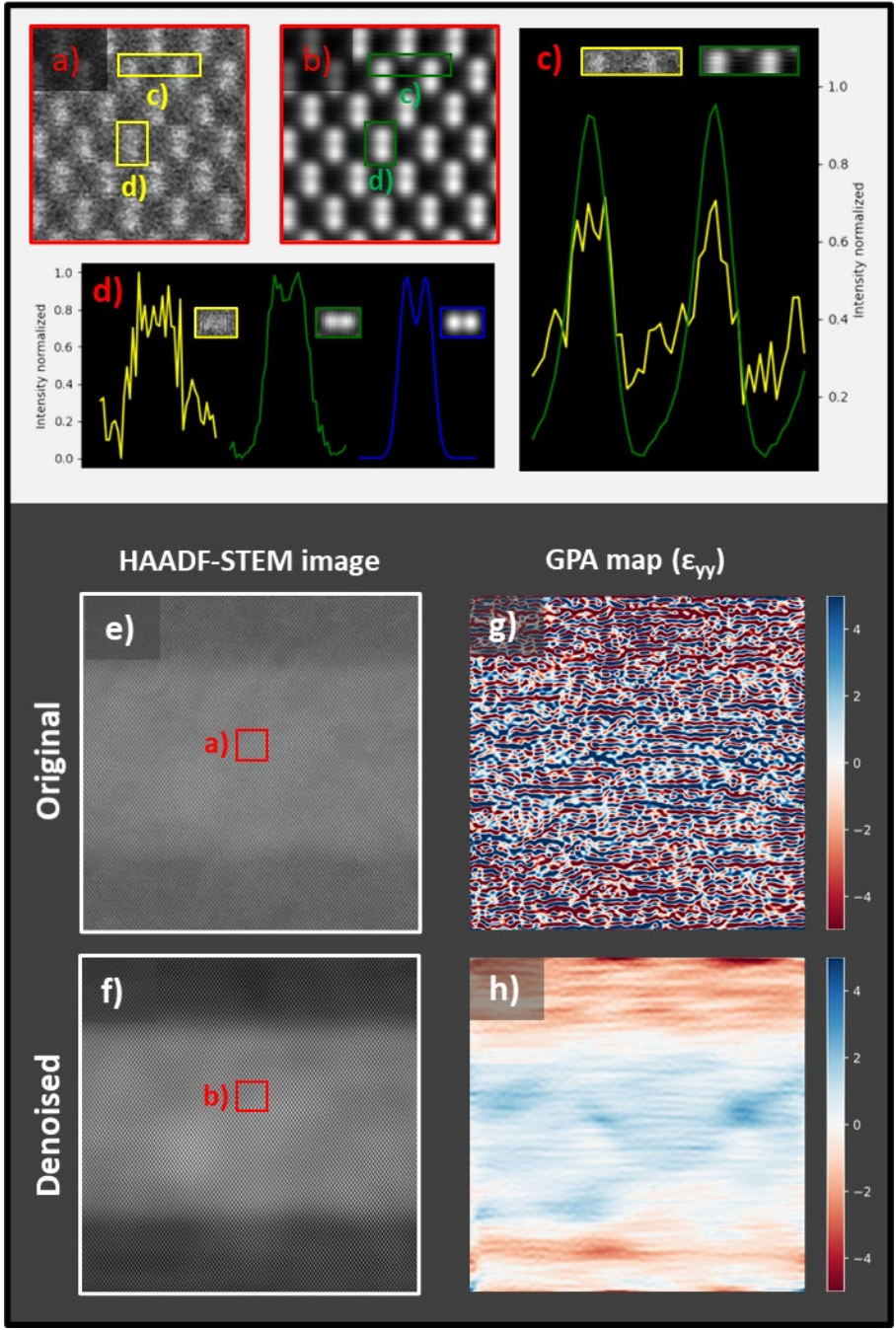

Figure 4: **Denoising results in a Ge QW SiGe heterostructure .** (a) Magnified detail extracted from the squared region in the HAADF STEM raw experimental image shown in (e). The region shows a detail obtained in the middle of the Ge QW, acquired along the [110] zone axis with a 1.86 nm field of view (FOV). (b) Corresponding denoised image. (c) Horizontal intensity profiles of Ge atomic columns in the raw (yellow) and denoised (green) images, showing a smoother distribution and an increased peak-to-valley contrast, indicating improved SNR. (d) Vertical intensity profiles of Ge dumbbells in the raw (yellow) and denoised (green) images, compared to the ideal theoretical profile (blue), demonstrating enhanced atomic resolution. The second part of the figure focuses on the impact of denoising on strain analysis: (e) HAADF-STEM image of a Ge QW cross-section acquired with a 38 nm FOV. (f) Denoised version of (e). The red squares indicate the selected regions magnified in panels (a) and (b). (g) and (h) Strain maps of the $\epsilon_{yy}$ component, calculated using the bottom SiGe barrier as a reference, for the raw and denoised images, respectively.

For the simulations, we used the abtem Python package [25]. All parameters used for both simulated images are shown in Table S3. After generating the ideal simulations (Fig.5g and h), Poisson noise with $\lambda$ = 0.4 was introduced to closely match the experimental conditions and render the sulfur positions indistinguishable (Fig.5j and k). Finally, we applied our denoising algorithm to obtain the denoised images for both $WS_2$ , with and without sulfur atoms. As shown in Fig.5m and n, only the simulation that originally contained S atoms shows intensity peaks at the expected sulfur positions, while the simulation without S does not. This confirms that our denoising model does not introduce artificial periodic structures but instead enhances real atomic features obscured by noise. Specifically, for $WS_2$ monolayers, our method enables the visualization of sulfur atoms, effectively surpassing the limitations imposed by the acquisition technique.

To further reinforce the validity of our method, we compared it against a recent CNN-based denoising model operating in real space (Figs.5b and 5c).[17] While both methods improve the overall clarity of HAADF-STEM images, only our frequency-domain approach reveals the low-intensity signals corresponding to sulfur atoms (Fig.5d and 5e). The real-space method fails to recover these features, underscoring the greater sensitivity of our frequency-based strategy to subtle structural information. This highlights the potential of our method for advancing the interpretation of electron microscopy data, particularly in cases involving low-Z elements in beam-sensitive materials.

# 4 Conclusions

In this work, we have introduced an automated deep learning-based denoising approach that operates in the frequency domain, leveraging a U-Net CNN trained on simulated FFT patterns. Unlike traditional filtering methods that rely on manual frequency selection, our model automatically identifies and enhances relevant frequency components, effectively suppressing noise while preserving critical crystallographic information. This method addresses a key limitation of conventional FFT filtering, which often introduces human bias and risks removing essential structural details.

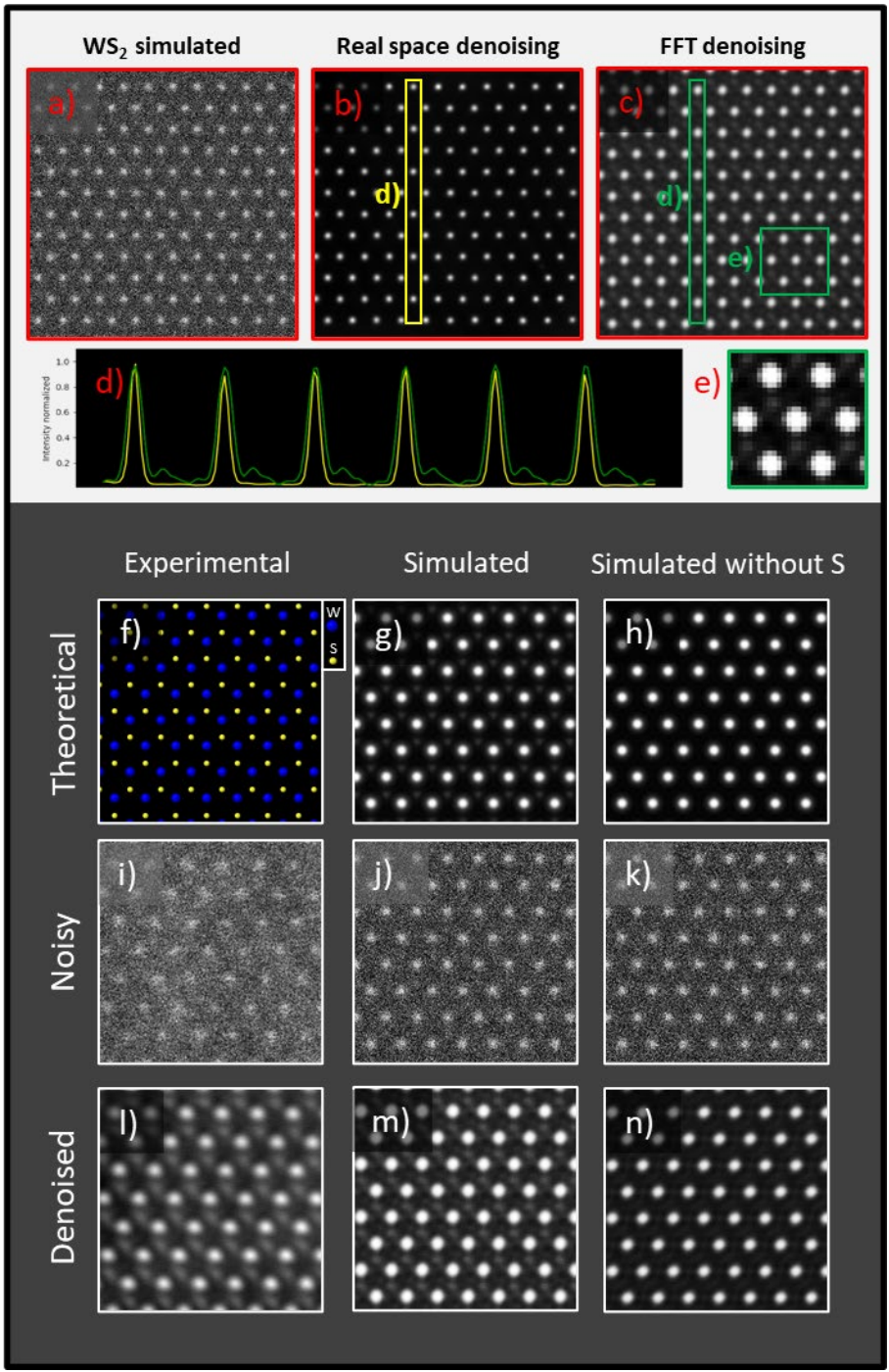

Figure 5: **Denoising results in a WS$_2$ monolayer.** The figure is divided into two sections. The top row shows denoising results for a simulated WS$_2$ monolayer. (a) Noisy HAADF-STEM simulation along the basal [0001] zone axis with a 20 nm field of view (FOV). (b) Result after applying a state-of-the-art real-space denoising algorithm [17] and (c) the result using our proposed frequency-domain denoising algorithm. (d) Vertical intensity profiles extracted from (b) and (c), highlighting W and S atoms in the real-space (yellow) and reciprocal-space (green) denoised images. (e) Zoomed-in region from (c), clearly showing the emergence of sulfur signals. The second part of the figure shows denoising results for a (i) HAADF-STEM image of an experimental WS$_2$ monolayer, acquired along the [0001] zone axis with a 10 nm FOV and the consequent simulation tests. (f) Theoretical atomic structure of hexagonal WS$_2$ in this orientation. (l) Denoised image from (i), where previously hidden sulfur positions become visible. To validate this effect, we perform simulated tests: (g) Simulated WS$_2$ structure including S atoms, and (h) a WS$_2$ simulation without S. (j) and (k) are the noisy versions of (g) and (h), mimicking experimental conditions. (m) and (n) show the corresponding denoised images of (j) and (k), confirming that the S atomic positions are retrieved only when originally present, ruling out artificial periodic artifacts from the algorithm.

Our approach has been validated on experimentally relevant material systems, demonstrating its ability to enhance fine structural details and reveal atomic features that would otherwise remain hidden by noise. By selectively amplifying meaningful frequencies, our model improves the visibility of low-contrast elements while maintaining spatial resolution, outperforming conventional denoising techniques. Despite these promising results, our method has certain limitations. Currently, it is designed for monocrystalline structures, and its performance in polycrystalline samples or regions with multiple grains is less reliable. In such cases, artifacts may appear at grain boundaries due to inconsistencies in FFT patterns across different crystallographic orientations. Addressing these challenges will be crucial for extending the applicability of this approach to a broader range of materials.

This study highlights the potential of deep learning for FFT-based denoising in (scanning) transmission electron microscopy atomic resolution images and opens new avenues for further exploration in this field. The ability to enhance frequency-domain information in an automated and data-driven manner paves the way for future advancements in denoising methodologies, potentially leading to more accurate and artifact-free image reconstructions.

# 5 Acknowledgments

ICN2 acknowledges funding from Generalitat de Catalunya 2021SGR00457. This study is part of the Advanced Materials programme and was supported by MCIN with funding from European Union NextGenerationEU (PRTR-C17.I1) and by Generalitat de Catalunya (In-CAEM Project). We acknowledge support from CSIC Interdisciplinary Thematic Platform (PTI+) on Quantum Technologies (PTI-QTEP+). This research work has been funded by the European Commission – NextGenerationEU (Regulation EU 2020/2094), through CSIC's Quantum Technologies Platform (QTEP). The authors thank support from the project AMaDE (PID2023-149158OB-C43), funded by MCIN/ AEI/10.13039/ /501100011033/ and by "ERDF A way of making Europe", by the "European Union". ICN2 is supported by the Severo Ochoa program from Spanish MCIN /

AEI (Grant No.: CEX2021-001214-S) and is funded by the CERCA Programme / Generalitat de Catalunya. Part of the present work has been performed in the framework of Universitat Autònoma de Barcelona Materials Science PhD program. JY has received funding from the CSC-UAB PhD scholarship program. IPH acknowledges funding from AGAUR-FI scholarship (2023FI-00268) Joan Oró of the Secretariat of Universities of the Generalitat of Catalonia and the European SocialPlus Fund. Authors acknowledge the use of instrumentation as well as the technical advice provided by the Joint Electron Microscopy Center at ALBA (JEMCA). ICN2 acknowledges funding from Grant IU16-014206 (METCAM-FIB) funded by the European Union through the European Regional Development Fund (ERDF), with the support of the Ministry of Research and Universities, Generalitat de Catalunya. ICN2 is founding member of e-DREAM. [26]